\documentstyle[12pt]{article}
\def\la{\mathrel{\mathpalette\fun <}}

\def\fun#1#2{\lower3.6pt\vbox{\baselineskip0pt\lineskip.9pt
\ialign{$\mathsurround=0pt#1\hfil##\hfil$\crcr#2\crcr\sim\crcr}}}
\author{N. Lehmann$^1$, D.V. Savin$^2$, V.V. Sokolov$^2$, H.-J. Sommers$^1$}
\title{Time Delay Correlations in Chaotic Scattering:\\Random Matrix
Approach}
\date{}

\begin{document}
\pagestyle{headings}
\maketitle

\begin{center}
$^1$Fachbereich Physik, Universit\"at Gesamthochschule Essen, D-4300 Essen,
Germany,\\
$^2$Budker Institute of Nuclear Physics, 630090 Novosibirsk, Russia

\end{center}

\begin{abstract}
We study the correlations of time delays in a model of chaotic resonance
scattering based on the random matrix approach. Analytical formulae  which
are valid for arbitrary number of open channels and arbitrary coupling
strength between resonances and channels are obtained by the supersymmetry
method. We demonstrate that the time delay correlation function, though being
not a Lorentzian, is characterized, similar to that of the scattering
matrix, by the gap between the cloud of complex poles of the $S$-matrix and
the real energy axis.
\end{abstract}

%\newpage

\section{Introduction}
The duration of particle collisions is an interesting and important aspect of
general scattering theory which is in a sense complementary  to the energy
representation ordinarily used. A collision is characterized in this case by
the time delay of particles in the region of interaction. Wigner \cite{W-55}
was the first who considered the time on which a monochromatic particle with
given
angular momentum is delayed during its elastic scattering. He established
the connection of this time delay to the energy derivative of the scattering
phase shift. The sharper the energy dependence of the phase shift is the
longer is the time delay.

Later on Smith \cite{S-60} extended the time delay concept on many channel
problems introducing the time delay matrix
\begin{equation}\label{q-matr}
 Q^{ab}(E) = -i\hbar\left\{\frac{d}{d\,\varepsilon}\,
 \sum_{c}S^{ac}(E\!+\!\frac{\varepsilon}{2})
 S^{\ast\,cb}(E\!-\!\frac{\varepsilon}{2})
 \right\}_{\varepsilon=0}\,\,,
\end{equation}
in the channel space. Here $S$ is the scattering matrix and the summation
index $c$ runs over all the $M$ open scattering channels. The matrix $Q$ is
hermitian; its diagonal element $Q^{cc}$ coincides with the mean duration of
collision (time delay) in the $c$-th entrance channel. Generally speaking,
the delays are different in different channels $c$. Taking the trace of the
Smith matrix, one arrives at the simple weighted-mean characteristic
\begin{equation}\label{q1}
Q(E) = \frac{1}{M}\,\sum_{c} Q^{cc}
     = -\frac{i}{M}\,\frac{d}{dE}\,\ln \det S(E)
\end{equation}
of the duration of collisions. Eq.  (\ref{q1}) is just the many-channel
version of the well-known simple Wigner formula. (Here and below we set
$\hbar=1$.)

The time delay turns out to be an especially pertinent concept for the chaotic
resonance scattering encountered in atomic, molecular and nuclear physics
\cite{L-77, LW-91}, as well as in the scattering of electromagnetic
microwaves \cite{Sr-91, GHLLRRSW-92, SS-92} in resonance billiard-like
cavities. The quantity $Q(E)$, being closely connected to the complex energy
spectrum of resonance states, shows in its energy dependence strong
fluctuations around a smooth regular variation. The two kinds of variation on
different energy scales are naturally decomposed
\begin{equation}\label{atd}
Q(E)=\langle Q(E)\rangle+Q_{fl}(E)\,\,,
\end{equation}
with an energy or ensemble averaging. By this, the slow energy dependence of
time delay is revealed whereas the two-point autocorrelation function
\begin{equation}\label{dcf}
C_Q(E,\varepsilon) =
\langle Q_{fl}(E\!+\!\frac{\varepsilon}{2})
Q_{fl}(E\!-\!\frac{\varepsilon}{2})\rangle=
\langle Q(E\!+\!\frac{\varepsilon}{2})
Q(E\!-\!\frac{\varepsilon}{2})\rangle -
\langle Q(E\!+\!\frac{\varepsilon}{2})\rangle
\langle Q(E\!-\!\frac{\varepsilon}{2})\rangle
\end{equation}
is used to characterize the time delay fluctuations.

To the best of our knowledge, the first consideration of these fluctuations
has been made numerically as well as analytically in \cite{WJ-89} and
\cite{SW-92} in the framework of rather peculiar model of resonance elastic
quantum scattering on a leaky surface of constant negative curvature. The
noteworthy feature of this model is that the poles of the scattering
amplitude turn out to correspond to zeros of the famous Riemann
$\zeta$-function. The real parts of the poles are therefore supposed
\cite{Mo-73} to be chaotically distributed similar to the eigenvalues of the
Gaussian Unitary Ensemble whereas all their imaginary parts (the widths of
resonances) are the same.  The latter very specific property partly deprives
the model of its interest since actual single-channel widths are known to
exhibit quite large fluctuations \cite{Po-65}.

The width fluctuations are suppressed when many channels are open. In this
case semiclassical approximation can be as a rule expected to be valid.  The
semiclassical analysis of the time delay in terms of closed periodic orbits
is given in \cite{Ec-93}. It is in particular emphasized there that only the
tail of the correlation function (\ref{dcf}) corresponding to the very large
values of $\varepsilon$ can immediately be related to the (short) periodic
orbits. Quite opposite, the central peak near the point $\varepsilon=0$ is
formed as a result of a strong interference of many orbits. Therefore, its
width describing the long-time  asymptotic behaviour of the Fourier transform
has no direct connection to the classical escape rate and has rather to be
calculated on the pure quantum ground. This is in line with the results of
the analysis \cite{GR-89} of distribution of the resonance widths in the
three discs scattering problem.

It is now generally acknowledged that the random matrix theory \cite{Me-67}
represents a suitable and reliable foundation for description of local
properties of dynamical quantum chaos \cite{BG-84}. We therefore use below a
random matrix model of chaotic scattering to calculate the time delay
autocorrelation function. We suppose as usual that the number $N$ of
resonances is asymptotically large and use the powerful supersymmetry
technique \cite{E-83}
first applied to chaotic scattering problems in \cite{VWZ-85}. The number $M$
of the (statistically equivalent) scattering channels
can be small or large or can even scale with the number of resonance states.
One can treat the latter two cases \cite{LW-91, HILSS-92, LSSS-94} as a
"semiclassical limit" in the matrix model. We show here that the time-delay
local fluctuations are governed, similar to those of the $S$-matrix
\cite{LSSS-94}, by the gap between the real axis and the upper edge of the
distribution of resonance energies in the complex energy plane. We compare
this result with that obtained in the framework of the periodic orbit
approach.

In the next section our statistical matrix model is briefly presented.
The connections of average time delay with the resonance spectrum and S-matrix
fluctuations are elucidated in sec.~3. After a short description in sec.~4
of the supersymmetry method which we use the main analytical results for the
time delay correlation function are given and discussed in detail in sec.~5.
Some numerical results shedding additional light upon properties of the time
delay correlations are gathered in sec. 6. We close with a brief summary
in sec.~7.

\section{The Resonance Matrix Model}
According to the general scattering theory, the evolution of the $N$-level
unstable system formed on intermediate stage of a resonance collision is
described \cite{MW-69, KNO-69, SZ-89} by the effective Hamiltonian
\begin{equation}\label{hamil}
{\cal H} = H - i\gamma\, W,\; \; \;\;  W = VV^{T}\,\,.
\end{equation}
The Hamiltonian (\ref{hamil}) acts within the intrinsic $N$-dimensional space
but acquires, due to the elimination of continuum variables, an antihermitian
part. The hermitian matrix $H$ is the internal Hamiltonian with a discrete
spectrum whereas the rectangular $N\times M$ matrix $V$ consists of the
amplitudes $V_m^c$ of transitions between $N$ internal and $M$ channel
states. These amplitudes are real in T-invariant theory, so that the matrix
$W$, similar to $H$, is real and symmetric. As usual, we neglect the smooth
energy dependence of $V$ and $W$.  The dimensionless parameter $\gamma$
characterizes the strength of the coupling of the internal motion to the
continuum.

The poles of the resonance scattering matrix in the complex energy plane
are those of the Green's function \cite{MW-69, KNO-69, SZ-89}
\begin{equation}\label{green}
{\cal G}(E) = (E-{\cal H})^{-1}\,\,.
\end{equation}
They coincide with the eigenvalues ${\cal E}_n=E_n-\frac{i}{2}\Gamma_n$ of
the effective Hamiltonian ${\cal H}$ with $E_n$ and $\Gamma_n$ being the
energy and width of $n$-th resonance state. It what follows, the properties
of the spectrum of complex energies ${\cal E}_n$ play the crucial role.

The intrinsic chaoticity of the internal motion of long-lived intermediate
system manifests itself by chaotic fluctuations in resonance scattering and
demands a statistical consideration. Therefore the random matrix approach
extending
the well-known \cite{Po-65, Me-67} description of chaotic bounded systems
has been worked out in \cite{W-84, VWZ-85, SZ-89}. It is usually
assumed that the hermitian part $H$ of the effective Hamiltonian belongs to
the Gaussian Orthogonal Ensemble (GOE),
\begin{equation}\label{goe}
\langle H_{nm} \rangle = 0,\ \ \ \langle H_{nm}H_{n'm'} \rangle =
\frac{\lambda^2}{N}(\delta_{nn'}\delta_{mm'}+\delta_{nm'}\delta_{mn'})\,\,.
\end{equation}
In the limit $N\rightarrow\infty$ eigenvalues of $H$ are situated in the
interval $[-2\lambda,2\lambda]$ with the density given by Wigner's
semicircle law. Following \cite{SZ-89}, we suggest the transition amplitudes
$V_n^c$ also to be statistically independent Gaussian variables,
\begin{equation}\label{rand}
\langle V^a_n \rangle = 0,\, \, \,
\langle V^a_nV^b_m \rangle = \frac{\lambda}{N}\delta^{ab}\delta_{nm}\,\,.
\end{equation}
We will use below the ensemble (\ref{goe},\ref{rand}) to calculate the
average quantities defined in (\ref{atd},\ref{dcf}).

\section{Time Delay and Resonance Spectrum}

Since we have neglected a smooth energy dependence of the effective
Hamiltonian (\ref{hamil}),
the poles ${\cal E}_n$ in the lower part of the complex energy
plane are the only singularities of the resonance scattering matrix.  Due to
the unitarity condition their complex conjugates ${\cal E}_n^*$ serve as
$S$-matrix's zeros. These two conditions result in the representation
\begin{equation}\label{det_s}
\det\,S(E) = \prod_{n} \frac{E-{\cal E}^{\ast}_{n}}{E-{\cal E}_{n}}\,\,.
\end{equation}
Substituting eq.(\ref{det_s}) in eq.(\ref{q1}), we come to the important
connection
\begin{equation}\label{q2}
 Q(E) = -2\,\mbox{Im}\,\left\{\frac{1}{M}\mbox{tr\,} {\cal G(E)}\right\}
 = \frac{1}{M}\,\sum_n \frac{\Gamma_n}{(E-E_n)^2+\frac{1}{4}\Gamma_n^2}
\end{equation}
between the time delay and the trace of the Green's function (\ref{green}) of
the intermediate unstable system. The time delay is entirely determined by
the spectrum of complex energies of this system. The collision duration
directly reflects the statistical properties of resonances. This is in
contrast to the scattering amplitudes $S^{cc'}$ which explicitly depend also
on the transition amplitudes $V_n^c$.

The ensemble averaging of eq.(\ref{q2}) gives
\begin{equation}\label{qav}
 \langle Q(E) \rangle = \frac{2}{m\lambda}\,\mbox{Re}\,{\sl g}(E)
\end{equation}
where $m < 1$ is the ratio $M/N$ and the function
\begin{equation}\label{gav}
 {\sl g}(E) = i\lambda\frac{1}{N}\,\langle \mbox{tr\,}\,{\cal G}(E)\rangle
\end{equation}
satisfies the cubic equation \cite{LSSS-94}
\begin{equation}\label{qeq}
 {\sl g}(E)-\frac{1}{{\sl g}(E)}+\frac{m\gamma}{1+\gamma{\sl g}(E)}-
 i\frac{E}{\lambda}=0\,\,.
\end{equation}
The (unique) solution with a positive real part has to be chosen.  It can be
seen from the consideration given in \cite{LSSS-94} that this real part is
close to $\frac{\lambda}{N}\pi\rho(E)$ with $\rho(E)$ being the projection on
the real energy axis near the scattering energy $E$ of the density of
resonance levels in the complex energy plane.

On the other hand, averaging eq.(\ref{q-matr}) directly, we express
$\langle Q \rangle$ in terms of the two-point $S$-matrix correlation function
\cite{VWZ-85, LSSS-94}. In the limit of a large number of statistically
equivalent channels, $M\gg 1$, scaling with the number of resonances $N$
\begin{equation}\label{q-ss}
\langle Q \rangle =
-i\frac{dC_S(\varepsilon)}{d\varepsilon}\Bigg|_{\varepsilon=0} +
i\frac{d{\cal T}(\varepsilon)}{d\varepsilon}\Bigg|_{\varepsilon=0}\,\,.
\end{equation}
Here \cite{LSSS-94}
\begin{equation}\label{scor}
 C_S(\varepsilon) = \frac{i\Gamma(\varepsilon)}{\varepsilon +
 i\Gamma(\varepsilon)}\,{\cal T}(\varepsilon) \equiv
 K(\varepsilon)\,{\cal T}(\varepsilon)
\end{equation}
with the two smooth functions defined by
\begin{equation}\label{G,T}
 \Gamma(\varepsilon) = \frac{m}{2}\lambda\,
 \frac{{\cal T}(\varepsilon)}{{\sl g}(\varepsilon/2)}\;\;\;,
 \;\;\;{\cal T}(\varepsilon) = \frac{4\gamma{\sl g}(\varepsilon/2)}
 {\left[1+\gamma{\sl g}(\varepsilon/2)\right]^2}
\end{equation}
and we set $E=0$ for the sake of simplicity. The quantity
\begin{equation}\label{trc}
C_S(0) = {\cal T}(0)\equiv T
\end{equation}
coincides with the transmission coefficient $T=1-|\langle S\rangle|^2$. With
eq.(\ref{scor}) taken into account we obtain from (\ref{q-ss})
\begin{equation}\label{qav2}
\langle Q \rangle =
-iT\frac{dK(\varepsilon)}{d\varepsilon}\Bigg|_{\varepsilon=0}
= \frac{T}{\Gamma_0}\,\,,
\end{equation}
where we have designated $\Gamma(0)$ as $\Gamma_0$.

As long as the typical values of the quantity $\Gamma(\varepsilon)$ are small
as compared to the parameter $\lambda$ characterizing the scale of the smooth
$\varepsilon$-dependence of the function ${\cal T}(\varepsilon)$, the two
factors on the r.h.s. of eq.(\ref{scor}) have quite different energy scales.
Only the first fast varying factor $K(\varepsilon)$ describes the local
fluctuations whereas the second one corresponds to the joint influence of all
resonances giving rise to the processes with a very short duration. The
latter came out from eq.(\ref{qav2}). The average time delay of a
non-monochromatic spatially small wave packet caused by the formation of a
long-lived intermediate state \cite{L-77, DHM-92} is determined just by the
factor $K(\varepsilon)$ \cite{LSSS-94}
\begin{equation}\label{dstd}
\langle \tau \rangle =
-i\frac{dK(\varepsilon)}{d\varepsilon}\Bigg|_{\varepsilon=0}
= \Gamma_0^{-1}\,\,.
\end{equation}
This implies the connection \cite{DHM-92, LSSS-94}
\begin{equation}\label{std}
\langle\tau\rangle = \langle Q\rangle/T =
\frac{2N}{\lambda MT}\,{\sl g}(0)\approx\frac{2\pi\rho}{MT}\,\,.
\end{equation}

\section{The Supersymmetry Method}

Now we calculate the correlation function (\ref{dcf}). Taking into account
the relation (\ref{q2}), one can cast eq.(\ref{dcf}) into the form
\begin{equation}\label{qq2}
C_Q(E,\varepsilon)= \frac{2}{M^2}\,\mbox{Re} \left\{
\langle \mbox{tr\,}{\cal G}(E\!+\!\frac{\varepsilon}{2})
\mbox{tr\,}{\cal G}^{\dagger}(E\!-\!\frac{\varepsilon}{2}) \rangle
- \langle \mbox{tr\,}{\cal G}(E\!+\!\frac{\varepsilon}{2}) \rangle
\langle \mbox{tr\,}{\cal G^{\dagger}}(E\!-\!\frac{\varepsilon}{2})
\rangle \right\}\,\,.
\end{equation}
We also define the normalized quantity
\begin{equation}\label{Nqq}
K_Q(E,\varepsilon)=\frac{C_Q(E,\varepsilon)}{{\langle Q(E)\rangle}^2}\,\,.
\end{equation}
The terms containing two Green's functions with poles at the same side from
the real energy axis are omitted in (\ref{qq2}). We will briefly return to
this point later.

In the limit $\gamma=0$, when the system gets closed, the correlation function
(\ref{qq2}) becomes proportional to the GOE density-density correlation which
consists \cite{Me-67} of the singular term $\delta(\pi\rho\varepsilon)$ and
Dyson's smooth function $-Y_2(\pi\rho\varepsilon)$. Coupling to the continuum
leads to appearing of a new energy scale caused by the decay processes. This
scale is defined \cite{LSSS-94} by the quantity $\Gamma(\varepsilon)$ from
eq.(\ref{G,T}).  One can anticipate a qualitative changing of the correlation
function to occur on this scale.  For larger distances the influence of the
antihermitian part should fade away and the asymptotics of $C_Q$ for
$\varepsilon\rightarrow\infty$ is expected to coincide with that of the
Dyson's function $-Y_2(\pi\rho\varepsilon)$.

To perform the ensemble averaging in (\ref{qq2}) we use the modification
worked out in \cite{LSSS-94} of the supersymmetry technique \cite{VWZ-85}.
Using the integral representation of Green's function as a multivariate
Gaussian integral over commuting and anticommuting variables, one gains the
possibility to accomplish the averaging exactly. With the help of the Fourier
transformation in the supermatrix space the integration over initial
auxiliary supervectors is then carried out. Going along this line, one finally
arrives at
\begin{equation}\label{grgr}
 \langle \mbox{tr\,}{\cal G}(E\!+\!\frac{\varepsilon}{2})\,
 \mbox{tr\,}{\cal G^{\dagger}}(E\!-\!\frac{\varepsilon}{2}) \rangle =
 -\frac{N^2}{4}\langle\mbox{str\,}(\sigma \eta_1)\,\mbox{str\,}
 (\sigma \eta_2) \rangle_{{\cal L}}\,\,.
\end{equation}
Here the shorthand $\langle\ldots\rangle_{{\cal L}}$ is used to denote
the integral
\begin{equation}\label{shand}
 \langle \ldots \rangle_{{\cal L}} =
 \int\!d[\sigma]\,d[\hat\sigma]\,\exp \{
 - N{\cal L}(\sigma,\hat\sigma) \} (\ldots)
\end{equation}
over two $8\times8$ supermatrices $\sigma$ and $\hat\sigma$ with the
measure defined by the Lagrangian \cite{LSSS-94}
\begin{equation}\label{Lg}
 {\cal L}(\sigma,\hat\sigma) = \frac{1}{4}\,\mbox{str\,}\sigma^2
 - \frac{i}{2}E\,\mbox{str\,}\sigma -
 \frac{i}{2}\mbox{str\,}(\sigma\hat\sigma) +
 \frac{1}{2}\mbox{str\,}\ln(\hat\sigma) + \frac{m}{2}\,\mbox{str\,} \ln
 (1\!+\!\gamma\sigma \eta) - \frac{i}{4}\varepsilon\,\mbox{str\,}(\sigma
 \eta)\,\,.  \end{equation} The  diagonal supermatrices appearing above are
equal to \[\eta=\mbox{diag}(1,1,-1,-1,1,1,-1,-1)\]
\[\eta_1=\mbox{diag}(1,1,0,0,-1,-1,0,0)\;\;\;
\eta_2=\mbox{diag}(0,0,1,1,0,0,-1,-1)\,\,. \]
Here we have set the GOE parameter $\lambda$ equal to one.

The supermatrix $\sigma$ can be decomposed in the following way \cite{VWZ-85}
\begin{equation}
\sigma=T_0\,\sigma_R\,T_0^{-1}
\end{equation}
where $T_0$ is a transformation from a non-compact manifold whereas the matrix
$\sigma_R$ is
diagonalized by transformations from a compact one. This implies a
corresponding decomposition of the integrals on the r.h.s. of (\ref{shand})
\begin{equation}\label{z4}
 \langle \ldots \rangle_{{\cal L}} =
 \int\!{\cal F}(\sigma_R)\,d[\sigma_R]\,d[\hat\sigma]
 \exp\{-N{\cal L}_R(\sigma_R,\hat\sigma) \}
 \int\!d\mu\exp\{-N{\cal L}_{\mu}(\sigma_R,T_0)\}(\ldots)\,\,.
\end{equation}
The Berezinian ${\cal F}(\sigma_R)$ depends only on the eigenvalues of
$\sigma_R$; $d\mu$ is the invariant measure of the manifold of non-compact
transformations $T_0$.  At last, the Lagrangian (\ref{Lg}) is splitted into
 two parts, ${\cal L}_R$ and ${\cal L}_{\mu}$, given by
\begin{equation}\label{dLg}
\begin{array}{l}
 {\cal L}_R(\sigma,\hat\sigma) = \frac{1}{4}\mbox{str\,}\sigma_R^2 -
 \frac{i}{2}E\,\mbox{str\,}\sigma_R -
 \frac{i}{2}\mbox{str\,}(\sigma_R\hat\sigma) +
 \frac{1}{2}\mbox{str\,}\ln(\hat\sigma)\,\,,  \\  \\ {\cal
 L}_{\mu}(\sigma_R,T_0) = -\frac{i}{4}\varepsilon\,\mbox{str\,}(\sigma_R
 T_0^{-1}\eta T_0) + \frac{m}{2}\,\mbox{str\,}\ln(1\!+\!\gamma\sigma_R
 T_0^{-1}\eta T_0)\,\,.  \end{array} \end{equation} Only the second part
${\cal L}_{\mu}$ depends on the non-compact variables. The first one
${\cal L}_R$ is invariant under a transformation by
$T_0$ since it is fully absorbed by an appropriate transformation of
$\hat\sigma$. One can easily verify that the corresponding Berezinian is
equal to unity.

Since the number of resonances $N\rightarrow \infty$, the integrations over
$\sigma_R$ and $\hat\sigma$ can be carried out in the saddle-point
approximation. At the same time, one has to integrate exactly over
non-compact variables as long as the number of channels $M$ is finite
($m=0$). The saddle-point approximation becomes valid for the latter
integration when the number $M$ also tends to infinity ($m$ is finite). We
will consider both cases mentioned. To simplify formulae we restrict
our further consideration to the center of the GOE spectrum $E=0$.

\section{Time Delay Correlation Function}

Let us first consider collisions with a fixed number of channels $M$. The
logarithmic term in ${\cal L}_{\mu}$ being proportional to the small ratio
$m$ does not influence then the saddle-point equations in the
$(\sigma_R,\hat\sigma)$-sector. In particular, the term in (\ref{qeq})
containing this ratio has to be omitted. The saddle-point equations are
trivially solved in this case and at the point $E=0$
\begin{equation}\label{sps1}
\hat\sigma=-i\sigma_R^{-1}\;\;,\;\;\sigma_R=\eta\,\,.
\end{equation}
With integrations over $\sigma_R$ and $\hat\sigma$ being done, the
correlation function (\ref{dcf}) reduces to the integral
\begin{equation}\label{Cqq}
 K_Q(\varepsilon)=2\,\mbox{Re}\int\!d\mu\,
 \mbox{str\,}(\kappa\alpha_1)\,\mbox{str\,}(\kappa\alpha_2)\,
 \exp\Bigl\{ \frac{i}{2}\pi\rho\varepsilon\,\mbox{str\,}\alpha_1
-\frac{M}{2}\mbox{str\,}\ln(1\!+\!\frac{1}{2}T\alpha_1) \Bigr\}
\end{equation}
over the invariant measure of the non-compact manifold of $T_0$-matrices.
Here $\alpha_{1,2}$ are the $4\times 4$ supermatrices defined in
\cite{VWZ-85}, the supermatrix $\kappa=\mbox{diag}(1,1,-1,-1)$ and
\begin{equation}\label{trc0}
T=\frac{4\gamma}{(1+\gamma)^2}
\end{equation}
is the transmission coefficient (\ref{trc}) calculated in the limit of $m=0$.

The further calculations go along the line described in details in
\cite{VWZ-85} and lead to the result
\[K_Q(\varepsilon)=\frac{1}{4}\,\int\limits_0^1\!d\lambda_0\!
\int\limits_0^{\infty}\!d\lambda_1\!\int\limits_0^{\infty}\!d\lambda_2\,
\mu(\lambda_0,\lambda_1,\lambda_2)(2\lambda_0+\!\lambda_1\!+\!\lambda_2\!)^2\,
\mbox{cos}\{\pi\rho\varepsilon(2\lambda_0+\!\lambda_1\!+\!\lambda_2\!)\}\]
\begin{equation}\label{K-f}
\times\left[ \frac{(1\!-\!T\lambda_0)^2}
{(1\!+\!T\lambda_1)(1\!+\!T\lambda_2) }\right]^{M/2}
\end{equation}
where
\[\mu(\lambda_0,\lambda_1,\lambda_2)=
\frac{(1\!-\!\lambda_0)\lambda_0|\lambda_1-\lambda_2|}
{[(1+\lambda_1)\lambda_1(1+\lambda_2)\lambda_2]^{1/2}
(\lambda_0\!+\!\lambda_1)^2(\lambda_0\!+\!\lambda_2)^2 }\,\,.\]

The dependence of the function $K_Q$ on openness of the unstable system is
fully contained in the last factor in (\ref{K-f}). If at least one of the
quantities $M$ or $T$ is equal to zero the threefold integral reduces to
the single one \cite{E-83}
\begin{equation}\label{d-d}
K_Q^{(0)}(\varepsilon)=\int\limits_0^2\!dt\,
t\left(1-\frac{1}{2}\ln(t\!+\!1)\right)
\mbox{cos}(\pi\rho\varepsilon t)+
\int\limits_2^{\infty}\!dt\,
\left(2-\frac{t}{2}\ln\frac{t\!+\!1}{t\!-\!1}\right)
\mbox{cos}(\pi\rho\varepsilon t)
\end{equation}
\[=\delta(\pi\rho\varepsilon)-Y_2(\pi\rho\varepsilon)\]
which is just the normalized GOE density-density correlation function.

Generally speaking, the threefold integral in (\ref{K-f}) can be investigated
for arbitrary number of channels $M$ only numerically using the methods
developed in \cite{V-86} (see the next section). However, this integral can
be simplified if $M$ becomes large enough. Let the number $M$ grow still
keeping the ratio $m=0$ and the product $MT=2\pi\rho\Gamma_W$ (compare with
(\ref{std})) fixed. The quantity $\Gamma_W$ is just the limiting value of
$\Gamma_0$ with $T$ and ${\sl g}$ calculated in the limit $m=0$. It coincides
with the well-known semiclassical Weisskopf estimate \cite{BW-79} of the
correlation length of Ericson fluctuations.  Then
\begin{equation}\label{t-exp}
\left[\frac{(1\!-\!T\lambda_0)^2}
{(1\!+\!T\lambda_1)(1\!+\!T\lambda_2)}\right]^{M/2}
\rightarrow\exp\{-\pi\rho\Gamma_W\,
(2\lambda_0+\!\lambda_1\!+\!\lambda_2\!)\},
\end{equation}
and one obtains similar to eq.(\ref{d-d})
\[K_Q(\varepsilon)=\int\limits_0^2\!dt\,t e^{(-\pi\rho\Gamma_W t)}\,
\left(1-\frac{1}{2}\ln(t\!+\!1)\right)
\mbox{cos}(\pi\rho\varepsilon t)\]
\begin{equation}\label{lM}
+\int\limits_2^{\infty}\!dt\,e^{(-\pi\rho\Gamma_W t)}\,
\left(2-\frac{t}{2}\ln\frac{t\!+\!1}{t\!-\!1}\right)
\mbox{cos}(\pi\rho\varepsilon t)\,\,.
\end{equation}
This is in close analogy with the consideration of the S-matrix correlation
function made in \cite{V-86}.

A new convergency factor appeared in the integrals in (\ref{lM}) as compared
to (\ref{d-d}) where only the oscillating cosine cuts the integral in the
region
of asymptotically large $t$. This makes the function $K_Q$ finite for all
values of $\varepsilon$ including zero, so that the $\delta$-function is now
smeared out. The behaviour of $K_Q(\varepsilon)$ is quite different in the
regions $\varepsilon\ll\Gamma_W$ and $\varepsilon\gg \Gamma_W$. In the first
one it is determined by decays and therefore is sensitive to the coupling to
the continuum. Quite opposite, for large $\varepsilon$ the behaviour becomes
universal since the GOE fluctuations described by the Dyson's function $Y_2$
are restored. It is perfectly reasonable since an open system cannot be
distinguished from a closed one during a small time $t\ll \Gamma_W^{-1}$.

The first $\gamma$-sensitive domain is widened when the width $\Gamma_W$
grows.
In the case of small $\rho\Gamma_W\ll 1$ (isolated resonances) it is natural
to set aside the contribution of asymptotics of the integrand presenting
(\ref{lM}) in the form
\begin{equation}\label{isol}
 K_Q(\varepsilon) = \frac{1}{\pi\rho}\,\frac{\Gamma_W}
{(\varepsilon^2\!+\!\Gamma_W^2)}\, +
\end{equation}
\[\int\limits_0^2\!dt\,e^{-\pi\rho\Gamma_W t}\,
\left(t-\frac{t}{2}\ln(t\!+\!1)-1\right)
\mbox{cos}(\pi\rho\varepsilon t) +
\int\limits_2^{\infty}\!dt\,e^{-\pi\rho\Gamma_W t}\,
\left(1-\frac{t}{2}\ln\frac{t\!+\!1}{t\!-\!1}\right)
\mbox{cos}(\pi\rho\varepsilon t)\,\,.\]
The Lorentzian contribution with the width $\Gamma_W$ directly traced to
the GOE $\delta$-function dominates in the domain $\varepsilon\la\Gamma_W$.
The sum of the integrals in the second line is negative for all values of
$\varepsilon$ and approaches asymptotically the function $Y_2$ from above.
We thus come to the conclusion that the correlation function vanishes at some
intermediate point $\varepsilon_0$ which can be estimated as
\begin{equation}\label{isoz}
\varepsilon_0\simeq \sqrt{\frac{\Gamma_W}{\pi\rho}}
\end{equation}
using the condition
\[\frac{1}{\pi\rho}\,\frac{\Gamma_W}{(\varepsilon_0^2\!+\!\Gamma_W^2)}
\sim |Y_2(\rho\varepsilon_0)|\sim 1\,\,.\]

The regime of strongly overlapping resonances, $\rho\Gamma_W\gg 1$, is the
most interesting. In this case the
main contribution in $K_Q$ comes from the region of small $t$. Therefore, the
second integral in (\ref{lM}) can be neglected. Dropping then out the small
logarithmic term in the first integral and extending its upper limit to
infinity, we arrive at
\begin{equation}\label{overl}
K_Q(\varepsilon)\approx\int\limits_0^\infty\!dt\,t e^{(-\pi\rho\Gamma_W t)}\,
\mbox{cos}(\pi\rho\varepsilon t) =
\frac{1}{\pi^2\rho^2}\frac{\Gamma_W^2-\varepsilon^2}
{(\varepsilon^2+\Gamma_W^2)^2}\,\,.
\end{equation}
Corrections to this result are of higher order with respect to the parameter
$(\rho\Gamma_W)^{-1}$. The function (\ref{overl}) is not a Lorentzian at all.
Decreasing quadratically in a small vicinity of the point $\varepsilon=0$, it
deviates subsequently from a Lorentzian, becomes zero at the point
$\varepsilon=\Gamma_W$, reaches a negative minimum and approaches at last
zero from below. Just the correlation function of such a form with $\Gamma_W$
substituted by the classical escape rate was conjectured in \cite{Ec-93}
as the limiting classical expression following from the periodic orbit
picture. However, there is no room for the classical escape rate in the
matrix models considered here.  One can see that the found form has in fact
quantum grounds.

One should return to the exact expressions (\ref{z4},\ref{dLg}) if the ratio
$m$ is finite. The resonances strongly overlap in this case. The
saddle-point is now found to be
\begin{equation}\label{sps2}
T_0=1\;\,,\;\, \hat\sigma=-i\sigma_R^{-1}\;\,,
\;\, \sigma_R={\sl g}(\varepsilon/2)\,\eta\,\,,
\end{equation}
where ${\sl g}$ is the  solution chosen in sec. 5 of the cubic equation
(\ref{qeq}).  The sequential saddle-point integrations over
$\sigma_R,\hat\sigma$ and then over the non-compact manifold result in the
expression
\begin{equation}\label{Cqqm}
K_Q(\varepsilon)=-\frac{4}{M^2T^2}\,\mbox{Re}\,
\frac{\Gamma_0^2}{\left[\varepsilon+i\Gamma(\varepsilon)\right]^2}
\end{equation}
where the function $\Gamma(\varepsilon)$ defined in (\ref{G,T}) is just the
one appearing when the $S$-matrix fluctuations are considered \cite{LSSS-94}.

The explicit dependence on $\varepsilon$ gives rise to a sharp variation of
the correlation function (\ref{Cqqm}) in the vicinity of zero if the typical
values $|\Gamma(\varepsilon)|\ll 1$ (see eq.(\ref{scor}) and the discussion
below). As long as the ratio $m$ is small, the quantity $\Gamma(\varepsilon)$
is small indeed and we can neglect its smooth $\varepsilon$-dependence for
all $\varepsilon\la\Gamma_0\approx\Gamma_W$.  Eq.(\ref{Cqqm}) is equivalent
to eq.~(\ref{overl}) within this domain. The asymptotic behaviour for large
$\varepsilon$ also does not change since $\Gamma(\varepsilon)$ remains
restricted for all $\varepsilon$. A small difference can appear only for
intermediate values of $\varepsilon$.

However, for larger values of $m$ the deviation can become noticeable even
near the point $\varepsilon=0$. In this case the next term in the power
expansion
\begin{equation}\label{pex}
\Gamma(\varepsilon)\approx\Gamma_0+\Gamma_0'\,\varepsilon
\end{equation}
with respect to the smooth $\varepsilon$-dependence should be taken into
account \cite{LSSS-94}. Because of the smoothness, the derivative $\Gamma_0'$
is small. One can see from eq.(\ref{qeq}) that this derivative is pure
imaginary.  The form (\ref{overl}) is now reproduced again for sufficiently
small $\varepsilon$ ,
\begin{equation}\label{Cqqmap}
K_Q(\varepsilon) = \frac{4\Gamma_g^2}{M^2T^2}\,
\frac{\Gamma_g^2-\varepsilon^2}{(\varepsilon^2+\Gamma_g^2)^2 }\,\,,
\end{equation}
with
\begin{equation}\label{Gg}
\Gamma_g=\frac{\Gamma_0}{1+i\Gamma_0'}\,\,.
\end{equation}
It has been proven in \cite{LSSS-94} that $\Gamma_g$, playing the role of the
correlation length of the Ericson fluctuations, coincides with the gap between
the distribution of resonance energies in the complex energy plane and the
real energy axis. Therefore we come to the conclusion that the properties of
fluctuations both of the $S$-matrix and time delay are described by the same
quantity, the gap $\Gamma_g$, rather than the classical escape rate.

Until now we neglected the "one-sided" contribution
\[\widetilde{C}_Q(\varepsilon)=\langle Q\rangle^2\,
\widetilde{K}_Q(\varepsilon)=\]
\begin{equation}\label{tqq2}
-\frac{2}{M^2}\,\mbox{Re}\left\{\langle\mbox{tr\,}{\cal G}(\frac{\varepsilon}
{2})\mbox{tr\,}{\cal G}(-\frac{\varepsilon}{2})\rangle
- \langle \mbox{tr\,}{\cal G}(\frac{\varepsilon}{2})\rangle
\langle\mbox{tr\,}{\cal G}(-\frac{\varepsilon}{2})
\rangle \right\}
\end{equation}
to the correlation function (\ref{dcf}). As long as $m=0$, this contribution
is of higher order in the parameter $N^{-1}$. However, this is not the case
when the ratio $M/N$ is finite. So one has to calculate (\ref{tqq2})
explicitly. The well-known replica method \cite{EA-75} turns out to be
sufficient for the latter purpose. Dropping here the corresponding rather
cumbersome
expressions we only note that the function $\widetilde{K}_Q(\varepsilon)$ is
entirely expressed in terms of the slowly varying ${\sl
g}(\frac{\varepsilon}{2})$ and varies slowly itself. It has got no pronounced
resonance behaviour around the point $\varepsilon=0$ and constitutes a
smooth background for the correlation function. Its value at the point
$\varepsilon=0$ is approximately equal to
\[\widetilde{K}_Q(0)\approx - \frac{1}{8N^2}\]
so that
\[\Bigg|\widetilde K_Q(0)/{K_Q(0)}\Bigg|
\approx\frac{1}{2}\,\left(\frac{\pi\rho\Gamma_0}{2N}\right)^2\,\,.\]
The ratio is small under the condition
\begin{equation}\label{con}
\pi\rho\Gamma_0\ll N \quad\mbox{or}\quad \Gamma_0\ll 1
\end{equation}
implying a clear-cut distinction of the local and global scales
\cite{LSSS-94}. Such a scale separation is necessary for matrix models to
be valid so far as the fluctuations are concerned.

The obtained form of the $\varepsilon$-dependence of the many-channel
correlation function $C_Q$ is close to that found in \cite{SW-92} for the
Gutzwiller's model of single-channel chaotic scattering on a space of
negative curvature. The same values of all resonance widths and the
outcoming possibility  for resonances to overlap are two specific features
of the model which are in fact in strong disagreement with properties of
the resonance spectra represented by matrix models. In particular, the
single-channel resonances cannot overlap at all in the latter models
\cite{SZ-89} and their widths fluctuate strongly. That is why our result
for $M=1$ (see below) differs noticeably from the correlation function
of ref.\cite{SW-92}. The situation changes when the number of channels is
large.  The width fluctuations diminish with the number $M$ of channels
growing. Since the time delay depends, according to (\ref{q2}), only on
properties of the complex energies of resonances and not on the number of
channels directly, the correlation functions become similar in the two
quite different cases compared.

It is worthy to note that the resonances overlapping strongly suppress the
time delay fluctuations. Indeed, eq.(\ref{isol}) gives for isolated
resonances
\[K_Q(0)=\frac{1}{\pi\rho\Gamma_W}\gg 1\]
whereas
\[K_Q(0)=\frac{1}{\pi^2\rho^2\Gamma_W^2}\ll 1\]
when they overlap. The duration of a collision thus becomes a good definite
quantity in the "quasiclassical" limit.

\section{Numerical results}
Excepting a few limiting cases considered above, further analytical study of
(\ref{K-f}) is not possible and one has to use numerical methods. However,
the threefold integral as it stands does not suit for numerical computation.
A very convenient substitution of the integration variables has been proposed
in \cite{V-86} to overcome all difficulties appearing. Following this author
we reduce the expression (\ref{K-f}) to the Fourier integral
\begin{equation}\label{fourier}
 K_Q(\varepsilon) =
 \int\limits_0^{\infty}\!dt\, F(t)\cos(\pi\rho\varepsilon t)
\end{equation}
with the Fourier transform $F(t)$ given by a double integral of a smooth
function quite convenient for the numerical work. The asymptotic behaviour
of $F(t)$ can be easily found explicitly
\begin{equation}\label{F-asymp}
 F(t) \sim \left\{ \begin{array}{ll} t
 &\ \ \mbox{for}\ t\ll 1 \\ (1+Tt)^{-M/2} &\ \
 \mbox{for}\ t\gg 1 \end{array} \right.\,\,.
\end{equation}

For a closed system ($T=0$) the Fourier transform $F(t)$ tends to unity in
the large-$t$ asymptotics. This results in the $\delta$-term in the GOE
density-density correlation. A singularity still survives even for an open
system with one or two decay channels. The asymptotics (\ref{F-asymp})
implies square root or logarithmic divergences correspondingly at the
point $\varepsilon=0$ in these two cases.

In Fig. 1 the function $K_Q(x)$ versus $x=\rho\varepsilon$ is plotted for the
case of a single open channel. The singular behaviour near zero as well as
GOE-like asymptotics are shown. The dashed line represents the Dyson's
function $-Y_2(\pi x)$. The calculation was made for the value $\gamma=1$;
only some small domain around zero is sensitive to the choice of $\gamma$.
The correlation function Fig.1 has little in common with that found in
\cite{SW-92}. This discrepancy is due to the strong fluctuations of
single-channel widths in our model in contrast to identical widths of all
resonances in Gutzwiller's one.

For $M>2$ the quantity $K_Q(0)$ is finite and the correlation function
approaches, as the number of channels grows, the asymptotics given by
(\ref{lM}). The Fig. 2 demonstrates this for the ratio
$K_Q(\varepsilon)/K_Q(0)$ in the
case of overlapping resonances. In asymptotic regime (\ref{overl}) such a
ratio is an universal function of the only variable $\varepsilon/\Gamma_W$.
One can see how the exact result (\ref{K-f}) gets more and more close to this
universal behaviour.

The Lorentzian peak should dominate the ratio
$K_Q(\varepsilon)/K_Q(0)$ in the domain
$\varepsilon/\Gamma_W\la (\pi\rho\Gamma_W)^{-\frac{1}{2}}\gg 1$ when
resonances are isolated (see (\ref{isoz})). Fig. 3 demonstrates this for two
values of coupling constant $\gamma$.

As it has been mentioned above, the function $K_Q(\varepsilon)$ vanishes at
some point $\varepsilon_0$.  The position of this point as the function of
the number of channels $M$ at several fixed values of $\gamma$ is shown in
Fig. 4 for three different values of the coupling constant $\gamma$. It is
clearly seen that the square root dependence for isolated resonances (see
(\ref{isoz})) is replaced by the linear one for overlapping ones.

%\begin{center}
%$<<<<<<<<<<$ the finite $m$ case $>>>>>>>>>>$
%\end{center}

\section{Summary.}
In this paper we have considered the fluctuations of the characteristic time
of collisions in the framework of a random matrix model of resonance chaotic
scattering.  These fluctuations are entirely due to the fluctuations of the
spectrum of complex resonance energies. We calculate analytically the time
delay correlation function and investigate its properties analytically and
numerically for different values of the number of channels and the strength
of the coupling to the continuum. For any values of these parameters this
function is far from being a Lorentzian. In particular, it vanishes at some
point which plays the role of the characteristic correlation length of the
fluctuations. In the "quasiclassical" limit of a large number of strongly
overlapping resonances this length is given, similar to that of the S-matrix
fluctuations, by the gap between the upper edge of the distribution of
complex energies of resonances and the real energy axis. We do not expect
that this quantity may be connected to the escape rate appearing in the
classical theory of chaotic scattering. The latter has been conjectured in
\cite{Sm-91} to be the semiclassical limit for the correlation length in
chaotic scattering.

\begin{center}
{\large\bf Acknowledgements}
\end{center}
We are grateful to F.Izrailev for his permanent interest to this work.
Financial support by the Deutsche Forschungsgemeinschaft through the SFB 237
is acknowledged. For two of us (V.V.S. and D.V.S.) the research described in
this publication was made possible in part by Grant No RB7000 from the
International Science Foundation.

%\newpage

%\newpage
\thispagestyle{empty}
\section*{Figures}
\begin{description}
\item[{\rm Fig. 1}]
   The time delay correlation function (\ref{K-f}) versus $x=\rho\varepsilon$
   for $M=1$ and $\gamma=1.0$. The dotted curve is the Dyson's function
   $-Y_2(\pi x)$.

\item[{\rm Fig. 2}]
   Overlapped resonances.
   The normalized function $K_Q(\varepsilon)/K_Q(0)$ versus
   $x=\varepsilon/\Gamma_W$ for  three values
   of $M=$5, 10 and 20 (dash-dotted, dashed and dotted curves) and
   $\gamma=1.0$ The solid curve is the asymptotic expression (\ref{overl}).

\item[{\rm Fig. 3}]
   Isolated resonances.
   The normalized correlation function (\ref{isol})
   for $\rho\Gamma_W=0.1$ and $\rho\Gamma_W=0.01$
   (dashed and dotted curves), and Lorentzian (solid curve).

\item[{\rm Fig. 4}]
   The zero $\rho\varepsilon_0$ of $K_Q(\varepsilon)$ as function of $M$
   for three coupling constants:
   $\gamma=0.01\,(\star)$ (in this case $\rho\varepsilon_0$ has been blown
   up by a factor 10), $\gamma=0.1\,(\bullet)$ and $\gamma=1.0\,(\circ)$.
   Solid and dashed lines are $\rho\Gamma_W$ and the dotted curve is the
   estimate (\ref{isoz}).

\end{description}

\end{document}